\documentclass[twocolumn,showpacs,preprintnumbers,amsmath,amssymb,aps]{revtex4}
\usepackage{mathrsfs}
\usepackage{epsf}
\def\alg#1{\mathscr #1}
\def\<{\langle}\def\>{\rangle}
\def\set#1{{\sf #1}}
 
\def\Tr{\operatorname{Tr}}

\def\Base{\set{B}}

\def\map#1{{\mathscr{#1}}}

\def\spc#1{\mathcal{#1}}

\def\Bndd#1,#2{\mathcal{B}(#1,#2)}

\def\Proof{\medskip\par\noindent{\bf Proof. }}

\def\Sex{S_\textrm{ex}}
\newtheorem{lemma}{Lemma}
\newtheorem{theo}{Theorem}
\newtheorem{corollary}{Corollary}

\begin{document}
\title{Inverting quantum decoherence by classical feedback from
  the environment} \author{Francesco Buscemi, Giulio Chiribella, and
  Giacomo Mauro D'Ariano} \affiliation{QUIT Group, Dipartimento di
  Fisica ``A.  Volta'', Universit\`a di Pavia, via A.  Bassi 6, I-27100
  Pavia, Italy} \homepage{http://www.qubit.it} \date{July 5, 2005}
\begin{abstract}
  We show that for qubits and qutrits it is always possible to
  perfectly recover quantum coherence by performing a measurement only
  on the environment, whereas for dimension $d>3$ there are situations
  where recovery is impossible, even with complete access to the
  environment. For qubits, the minimal amount of classical information to
  be extracted from the environment equals the {\em entropy exchange}.
\end{abstract}
\maketitle

Decoherence is universally considered, on one side, as the major
practical limitation for communication and processing of quantum
information. On the other side, decoherence yields the
key concept to explain the transition from quantum to classical
world\cite{decoherence} due to the uncontrolled and unavoidable
interactions with the environment. Great effort in the literature has
been devoted to combat the effect of decoherence by engineering robust
encoding-decoding schemes: by storing quantum information on a larger
Hilbert space---as in quantum error correction\cite{ErrorCorr}---or by
protecting the quantum information in a decoherence-free
subspace\cite{DecoFree}, or else via topological
constraints\cite{TopoCostr}. Moreover, some authors have recently
addressed a different approach to undo quantum noises by extracting
classical information from the environment\cite{GW} and exploiting it
as an additional amount of side information useful to improve quantum
communication performances\cite{capacity}.

The recovery of quantum coherence from the environment is often a
difficult task, e. g.  when the environment is ``too big'' to be
controlled, as for spontaneous emission of radiation\cite{Plenio}.  By
regaining control on the environment the recovery can sometimes be
actually accomplished, for example by keeping the emitted radiation
inside a cavity. However, in some cases, the full recovery of quantum
coherence becomes impossible even in principle, namely even when one
has complete access to the environment. This naturally leads us to
pose the following question: in which physical situations is possible
to perfectly recover quantum coherence by monitoring the environment?

In this Letter we will show that for qubits and qutrits it is always
possible to perfectly cancel the effect of decoherence by
monitoring---i.~e. measuring---the environment. On the contrary, for
quantum systems with larger dimension $d$, namely for qudits with
$d>3$, there are situations where the recovery is impossible even in
principle. In order to prove the above assertion we will give a
complete classification of decoherence maps for any finite dimension
$d$, showing that they are always of the form of a Schur map. For
qubits we will also evaluate the minimal amount of classical
information that must be extracted from the environment in order to
invert the decoherence process.

A completely decohering evolution asymptotically cancels any quantum
superposition when reaching the stationary state, making any state
diagonal in some fixed orthonormal basis---the basis depending on the
particular system-environment interaction.  In the Heisenberg picture
we say that such a completely decohering evolution asymptotically maps
the whole algebra of quantum observable into a ``maximal classical
algebra'', that is a maximal set of commuting---namely jointly
measureable---observables.  It is also possible to consider partial
decoherence, i.~e. preserving superpositions within subspaces of the
total system, reducing the initial state into a block-diagonal form.
Here, however, we focus our attention on the worst case scenario of
complete decoherence, and also show how the present results can be
easily generalized to partial decoherence.

Let's denote by $\alg A_q$ the ``quantum algebra'' of all bounded
operators on the finite dimensional Hilbert space $\spc H$, and by
$\alg A_c$ the ``classical algebra'', namely any maximal Abelian
subalgebra $\alg A_c\subset\alg A_q$.  Clearly, all operators in $\alg
A_c$ can be jointly diagonalized on a common orthonormal basis, which
in the following will be denoted as $\Base =\{|k\>~|~k=1, \dots ,d\}$.
Then, the classical algebra $\alg A_c$ is also the linear span of the
one-dimensional projectors $|k\>\<k|$, whence $\alg A_c$ is a
$d$-dimensional vector space. According to the above general
framework, we call \emph{(complete) decoherence map} a completely
positive identity-preserving (i. e.  trace-preserving in the
Schr\"odinger picture) map $\map E$ which asymptotically maps any
observable $O \in \alg A_q$ to a corresponding ``classical
observable'' in $\alg A_c$, namely such that the limit $\lim_{n\to
  \infty} \map E^n (O)$ exists and belongs to the classical algebra
$\alg A_c$ for any $O \in \alg A_q$. Here we denote with $\map E^n$
the $n$-th iteration of the map $\map E$, implicitly assuming
markovian evolution.

It is easy to see that the set of decoherence maps is convex (i. e. if
we mix two decoherence maps we obtain again a decoherence map). Such
convex set will be denoted by $\set D$. Moreover, $\set D$ is a subset
of the convex set of maps that preserve all elements of the classical
algebra $\alg A_c$. Such maps have a remarkably simple form:
\begin{theo}[Schur form]\label{GeneralForm} A map $\map E$ preserves
  all elements of the maximal classical algebra if and only if it has
  the form
\begin{equation}\label{SchurMap}
\map E(O)= \xi \circ O.
\end{equation}
$A \circ B$ denoting the Schur product of operators $A$ and $B$, i.~e.
$A \circ B \equiv \sum_{k,l=1}^d A_{kl} B_{kl} |k\>\<l|$, $\{A_{kl}\}$
and $\{B_{kl}\}$ being the matrix elements of $A$ and $B$ in the basis
$\Base$, and $\xi_{kl}$ being a correlation matrix, i.e. a positive
semidefinite matrix with $\xi_{kk}=1$ for all $k=1, \dots ,d$.
\end{theo}   
\Proof Consider a Kraus representation of the map $\map E$:
\begin{equation}\label{Kraus}
\map E(O)= \sum_{i=1}^r~ E_i^{\dag}OE_i.
\end{equation} 
Exploiting a result by Lindblad\cite{Lind}, we know that a map $\map
E$ preserves all elements of an algebra $\alg A$ if and only if its
Kraus operators commute with the algebra itself, i.~e. $[E_i,O]=0$ for
any $O \in \alg A$.  Since in our case the algebra is the maximal
Abelian algebra $\alg A_c$, such a commutation relation implies $E_i
\in \alg A_c$, therefore
\begin{equation}\label{KrausOp}
E_i=\sum_{k=1}^{d}~e_k^{(i)}~|k\>\<k|.
\end{equation}
Substituting Eq. (\ref{KrausOp}) into Eq. (\ref{Kraus}), we obtain
\begin{equation}\label{SchurInProof}
\map E (O)= \sum_{k,l=1}^d~ \xi_{kl}~O_{kl}~|k\>\<l|,
\end{equation}
where 
\begin{equation}\label{XiDef}
\xi_{kl}\equiv \sum_{i=1}^r~ e^{(i)*}_k e^{(i)}_l.
\end{equation}
By definition, the matrix $\xi_{kl}$ is positive semidefinite, and the
identity-preserving condition $\map E (\openone)=\openone$~ in Eq.
(\ref{SchurInProof}) gives $\xi_{kk}=1$ for all $k$. 
Viceversa, it is obvious to see that any map of the form
(\ref{SchurMap}) preserves all elements of the classical algebra.
$\blacksquare$\medskip

In the case of partial decoherence, the Shur form (\ref{SchurMap}) generalizes to $\map
E (O)= \sum_{k,l}\xi_{kl}P_kOP_l$, where $P_k$'s are the
orthogonal projections over the invariant subspaces.

Since there is a linear correspondence between maps preserving $\alg
A_c$ and correlation matrices, the two sets share the same convex
structure, whence the map is extremal if and only if its correlation
matrix is extremal.

Thanks to Theorem \ref{GeneralForm} it is immediate to recognize the
general form of a decoherence map:
\begin{corollary} A map $\map E$ is a decoherence map if and only if
  it has the form (\ref{SchurMap}) where $\xi_{kl}$ is a correlation
  matrix with $|\xi_{kl}|<1$ for all $k\neq l$.
\end{corollary}
Notice that positivity of $\xi$ implies $|\xi_{kl}|\leq 1$, while
the requirement that $\lim_{n\to\infty}\map E^n(\alg A_q)=\alg A_c$
needs $|\xi_{kl}|< 1$ strictly.  This also implies the following:
\begin{corollary}
  The closure $\overline{\set D}$ of the set $\set D$ of decoherence
  maps coincides with the set of maps that preserve the classical
  algebra.
\end{corollary}
As examples of maps on the border of $\overline{\set D}$, simply
consider the identity map, or the map $\map U(\cdot)=U^\dag\cdot U$,
with the unitary $U$ diagonal on the basis $\Base$.

Another relevant property of the decoherence maps is the following:
\begin{corollary}\label{Corollary3}
  All decoherence maps commute among themselves, i.~e. their order is
  irrelevant.
\end{corollary}

The action of a decoherence map on quantum states is given in
Schr\"odinger picture by
\begin{equation}\label{SchurFormSchro}
\map E_S(\rho)= \xi^T \circ \rho,
\end{equation}
where $T$ denotes transposition with respect to the basis $\Base$
(also $\xi^T$ is a correlation matrix).  As a consequence, one has
exponential decay of the off-diagonal elements of $\rho$, since
$|~[\map E_S^n (\rho)]_{kl}~|=|\xi_{lk}|^n\cdot |\rho_{kl}|$. In other
words, any initial state $\rho$ decays exponentially towards the
completely decohered state $\sum_k~
\rho_{kk}~|k\>\<k|\equiv\rho_\infty$.
\begin{lemma}\label{l:extr} 
  A map $\map E$ is extremal in $\overline{\set D}$ if and only if it
  is extremal in the set of all maps.
\end{lemma}
\Proof Take $\map E$ extremal in $\overline{\set D}$, and suppose by
contradiction that in the set of all maps there are two maps $\map
E_1$ and $\map E_2$ such that $\map E=p \map E_1 +(1-p) \map E_2$.
Since $\map E$ leaves all elements of $\alg A_c$ invariant, for all
$k$ one has
\begin{equation}
|k\>\<k|=\map E (|k\>\<k|)=p \map E_1 (|k\>\<k|) +(1-p) \map E_2
(|k\>\<k|).
\end{equation} 
But $\map E_1$ and $\map E_2$ are positive and identity-preserving
maps, whence necessarily $\map E_i (|k\>\<k|)=|k\>\<k|$ for $i=1,2 $
and for all $k$, namely $\map E_1$ and $\map E_2$ are both in
$\overline{\set D}$. But $\map E$ is extremal in $\overline{\set D}$,
whence $\map E_1= \map E_2=\map E$, and $\map E$ is extremal in the
set of all maps. The converse direction is trivial.
$\blacksquare$\medskip

As a consequence of Lemma \ref{l:extr}, the convex structure of
decoherence maps can be obtained by application of the well known Choi
Theorem\cite{Choi}, which states that the canonical Kraus operators
$\{E_i\}$, $1\leq i \leq r$, of every extremal map are such that their
products $\{E_i^{\dag}E_j\}$, $1\leq i,j\leq r$, are linearly
independent.  A relevant consequence of this characterization is the
following
\begin{theo}
  If $\map E\in\overline{\set D}$ is extremal, then $r \leq \sqrt d$.
  For qubits and qutrits any map in $\set D$ is random-unitary.
\end{theo}
\Proof Due to linear independence, the dimension of the linear span of
$\{E_i^{\dag}E_j\}$ must be $r^2$. But this set of operators is a
subset of $\alg A_c$, whose dimension is $d$. Therefore $r^2\leq d$.
In particular, for $d \leq3$ one necessarily has $r=1$, i. e. the
extremal points of $\overline{\set D}$ are unitary maps.
$\blacksquare$\medskip

This means that for qubits and qutrits every decoherence map can be
written as
\begin{equation}\label{RandomUnitary}
\map E(O)=\sum_ip_iU_i^\dag OU_i,
\end{equation}
for some commuting unitary operators $U_i\in\alg A_c$ and probability
distribution $p_i$. Now, in Ref.\cite{GW} it is shown that the only
channels that can be perfectly inverted by monitoring the environment
are the random unitary ones. Therefore, it follows that one can
perfectly correct any decoherence map for qubits and qutrits by
monitoring the environment. The correction is achieved by retrieving
the index $i$ in Eq. (\ref{RandomUnitary}) via a measurement on the
environment, and then by applying the inverse of the unitary
transformation $U_i$ on the system (for pure joint system-environment
states the unitary form of the conditioned system transformations also
follows from the Lo-Popescu Theorem\cite{LoPop} for LOCC
transformations).  Therefore, the random-unitary map simply leaks
$H(p_i)$ bits of classical information into the environment ($H$
denoting the Shannon entropy), and the effects of decoherence can be
completely eliminated by recovering such classical information,
without any prior knowledge about the input state.

The fact that decoherence maps are necessarily random-unitary is true
only for qubits and qutrits.  Indeed, for dimension $d\geq 4$ there are
decoherence maps which are not random-unitary, since there exist
extremal correlation matrices whose rank is greater than
one\cite{LiTam}, e.~g. the rank-two matrix
\begin{equation}\label{Example}
\xi= \left(
\begin{array}{cccc}
 1 & 0 & \frac{1}{\sqrt{2}} & \frac{1}{\sqrt{2}} \\ 0 & 1 &
 \frac{1}{\sqrt{2}} & \frac{i}{\sqrt{2}}\\ \frac{1}{\sqrt{2}} &
 \frac{1}{\sqrt{2}} & 1 & \frac{1+i}{2} \\
\frac{1}{\sqrt{2}} & \frac{-i}{\sqrt{2}}& \frac{1-i}{2} & 1
\end{array}
\right).
\end{equation}
The canonical Kraus decomposition of the map $\map E (O)=\xi \circ O$,
can be obtained by diagonalizing the operator $\xi$ as
$\xi=|v_1\>\<v_1|+|v_2\>\<v_2|$, with $\<v_1|v_2\>=0$. Then, the the
canonical Kraus operators are $E_i= \sum_k\<k|v_i\>|k\>\<k|$, $i=1,2$,
and the corresponding map $\map E=\sum_i E_i^\dag\cdot E_i$ is not
unitary (its canonical Kraus decomposition contains two terms), nor is
random-unitary, since it is extremal.  Such decoherence maps with
$r\geq2 $ represent a process which is fundamentally different from
the random unitary one, corresponding to a {\em leak of quantum
  information} from the system to the environment, information that
cannot be perfectly recovered from the environment\cite{GW}.

Now we address the problem of estimating the amount of classical
information needed in order to invert a random-unitary decoherence
map.  If the environment is initially in a pure state, say $|0\>_e$, a
useful quantity to deal with is the so-called entropy
exchange\cite{Schumacher} $\Sex$ defined as
\begin{equation}\label{DefSex}
  \Sex(\rho)=S(\sigma_{e}^\rho),
\end{equation}
where $\sigma_{e}^\rho$ is the reduced environment state after the
interaction with the system in the state $\rho$, and
\mbox{$S(\rho)=-\Tr[\rho\log\rho]$} is the von Neumann entropy.  In
the case of initially pure environment, the entropy exchange depends
only on the map $\map E$ and on the input state of the system $\rho$,
regardless of the particular system-environment interaction chosen to
model $\map E$.  It quantifies the information flow from the system to
the environment and, for all input states $\rho$, one has the
bound\cite{Schumacher} $|S(\map E_S(\rho))-S(\rho)|\leq \Sex(\rho)$,
namely the entropy exchange $\Sex$ bounds the entropy production at
each step of the decoherence process.

In order to explicitly evaluate the entropy exchange for a decoherence
process, we can then exploit a particular model interaction between
system and environment. This can be done starting from Eq.
(\ref{XiDef}) noticing that it is always possible to write
$\xi_{kl}=\<e_k|e_l\>$ for a suitable set of normalized vectors
$\{|e_k\>\}$.  Then, the map $\map E_S(\rho)=\xi^T\circ\rho$ can be
realized as \mbox{$\map E_S(\rho)=\Tr_e[U(\rho\otimes
  |0\>\<0|_e)U^\dag]$}, where the unitary interaction $U$ gives the
transformation
\begin{equation}\label{Unitary}
U|k\>\otimes|0\>_e=|k\>\otimes|e_k\>.
\end{equation}
The final reduced state of the environment is then
$\sigma_{e}^\rho=\sum_k\rho_{kk}|e_k\>\<e_k|$.  Then, in order to
evaluate $\Sex$ for a decoherence map $\map E_S(\rho)=\xi^T\circ\rho$,
it is possible to bypass the evaluation of the states $|e_i\>$ of the
environment, using the formula
\begin{equation}\label{Sex_and_xi}
  \Sex(\rho)=S(\sqrt{\rho_\infty}\xi\sqrt{\rho_\infty}),
\end{equation}
which follows immediately from the fact that
$\sqrt{\rho_\infty}\xi\sqrt{\rho_\infty}$, and $\sigma_{e}^\rho$ are
both reduced states of the same bipartite pure state
$\sum_i\sqrt{\rho_{ii}}|i\>|e_i\>$.

The unitary interaction $U$ in Eq. (\ref{Unitary}) generalizes the
usual form considered for quantum measurements\cite{vonNeumann}, with
the quantum system interacting with a pointer, which is left in one of
the (nonorthogonal) states $\{|e_k\>\}$. The more the pointer states
are ``classical''---i.~e. distinguishable---the larger is the entropy
exchange, whence the faster is the decoherence process. In the limit
of orthogonal states, decoherence is istantaneous, i.~e. $\map
E_S(\rho)=\rho_\infty$. If the state of the system is a pure classical
one $\rho=|j\>\<j|$, the entropy exchange is zero, since
$\sigma_e^\rho=|e_j\>\<e_j|$ is pure.  In this case the environment
evolves freely, with the system untouched. For mixed classical state
$\rho$ there is a nonvanishing entropy flow, even if the state of the
system doesn't change.  This is because the entropy flow is well
defined only for a closed system---i.~e.  described by a unitarily
evolving global pure state, with \mbox{$\Delta
  S_\textrm{tot}=0$}---whence in the entropic balance one must
consider also a reference system $r$ purifying $\rho$. As an example,
let $\alg A_c\ni\rho=\sum_ip_i|i\>\<i|$ be purified as
$|\Psi\>=\sum_i\sqrt{p_i}|i\>_r|i\>$. Then, the action of
$\openone_r\otimes U$ on $|\Psi\>|0\>_e$ is $(\openone_r\otimes
U)|\Psi\>|0\>_e= \sum_i\sqrt{p_i}|i\>_r|i\>|e_j\>$ and the reduced
reference+system state changes according to
\begin{equation}
|\Psi\>\<\Psi|\longmapsto R=\sum_{ij}\xi_{ji}\sqrt{p_ip_j}|i\>\<j|_r\otimes|i\>\<j|,
\end{equation}
corresponding to $\map E_S(\rho)=\Tr_r[R]\equiv\rho$. In other words,
the non-null entropy exchange results in a decrease of the
correlations between the reference and the system.

When a map can be inverted by monitoring the environment---i.~e. in
the random-unitary case---the entropy exchange $\Sex(\openone/d)$
provides a lower bound to the amount of classical information that
must be collected from the environment in order to perform the
correction scheme of Ref.\cite{GW}. In fact, assuming a random-unitary
decomposition (\ref{RandomUnitary}) and using the
formula\cite{Schumacher} $\Sex (\rho)=S\left( \sum_{i,j} \sqrt{p_i
    p_j} \Tr[U_i \rho U_j^{\dag}] |i\>\<j| \right)$, we obtain
\begin{equation}\label{SexAndEntropy}
\Sex(\openone/d) \leq H(p_i)~.
\end{equation} 
The inequality comes from the fact that the diagonal entries of a
density matrix are always majorized by its eigenvalues, and it becomes
equality if and only if $\Tr[U_iU_j^{\dag}]/d=\delta_{ij}$, i.~e. the
map admits a random-unitary decomposition with \emph{orthogonal}
unitary operators. Moreover, from Eq. (\ref{Sex_and_xi}) we have $\Sex
(\openone/d)=S(\xi/d)$.

\emph{For qubits, $S(\xi/d)$ quantifies exactly the
  minimum amount of classical information} which must be extracted
  from the environment. Indeed, a unitary map in $\overline{\set D}$
  corresponds to the rank-one correlation matrix $\xi=|\phi\>\<\phi|$,
  where 
\begin{equation}\label{EqVec}
|\phi\>=\sum_{k=1}^d ~e^{i\phi_k}|k\>~.
\end{equation}  
For $d=2$ it is simple to see that any correlation matrix $\xi$ can be
diagonalized using two such vectors, i.~e. $\xi=p_1
|\phi_1\>\<\phi_1|+p_2|\phi_2\>\<\phi_2|$, whence the corresponding
map is random-unitary with $U_i=\sum_k \<k|\phi_i\>|k\>\<k|$. Clearly,
$\<\phi_1|\phi_2\>=0$ implies $\Tr[U_1^\dag U_2]=0$. Therefore for
qubits $H(p_i)=S(\xi/d)$.

Notice that the same decoherence map may be obtainable by a
random-unitary transformation with more than two outcomes, and a
flattened probability distribution $\{p_i\}$, corresponding to a
larger information $H(p_i)$. However, for qubits it is always possible
to perform a suitable measurement on the environment and to invert the
decoherence map retrieving the minimal amount of information from the
environment.

For dimension $d>2$, the bound in Eq. (\ref{SexAndEntropy}) is
generally strict.  Already for dimension $d=3$, even if all
decoherence maps are random-unitary, the amount of information
required for perfect correction may exceed $S(\xi/d)=\Sex(\openone/d)$.
As an example, the correlation matrix with nondegenerate spectrum
\begin{equation}
\xi=\begin{pmatrix}
1 & 0 & 1/\sqrt{2}\\
0 & 1 & 1/\sqrt{2}\\
1/\sqrt{2} & 1/\sqrt{2} & 1
\end{pmatrix}~.
\end{equation}
has the eigenvector $|v\>=|1\>-|2\>$, which is not of the form
(\ref{EqVec}). This means that it is not possible to write $\map E_S
(\rho)=\xi^T \circ \rho$ as a convex combination of orthogonal
unitaries.

Finally, it is worth noticing that, when a decoherence process can be
inverted, this can be done regardless of the number of iterations of
the map, since the iterated map is also random-unitary.  Clearly one
needs to perform the measurement on a larger Hilbert space for the
environment, however, the complexity of the measurement does not
necessarily increase. In fact, in order to restore the initial state
we only need to know how many times the unitary $U_i$ for each $i$ has
been applied to the system, since the unitary operators for different
$i$ commute and their order is irrelevant.

In summary, in this Letter we showed that for qubits and qutrits it is
always possible to perfectly invert decoherence by extracting
classical information from the environment. For dimension $d=4$,
instead, we gave a counterexample proving that for $d>3$ generally the
recovery is impossible even with complete access to the environment. A
complete classification of decoherence maps for any finite dimension
$d$ has been given in form of a Schur product with a correlation
matrix $\xi$.  The minimal amount of classical information needed to
invert decoherence has been evaluated for qubits as the von Neumann
entropy of $\xi/d$.

F~B and G~C acknowledge stimulating discussions with Mario~Ziman.
This work has been co-founded by the EC under the program ATESIT
(Contract No. IST-2000-29681), and the MIUR cofinanziamento 2003.

\end{document}